\documentclass[english,prd]{revtex4}
\usepackage{babel}

\usepackage{graphicx}
\usepackage{amssymb}
\usepackage{esint}
\usepackage[unicode=true, pdfusetitle,
 bookmarks=true,bookmarksnumbered=false,bookmarksopen=false,
 breaklinks=false,pdfborder={0 0 1},backref=false,colorlinks=false]
 {hyperref}

\makeatletter
\@ifundefined{textcolor}{}
{%
 \definecolor{BLACK}{gray}{0}
 \definecolor{WHITE}{gray}{1}
 \definecolor{RED}{rgb}{1,0,0}
 \definecolor{GREEN}{rgb}{0,1,0}
 \definecolor{BLUE}{rgb}{0,0,1}
 \definecolor{CYAN}{cmyk}{1,0,0,0}
 \definecolor{MAGENTA}{cmyk}{0,1,0,0}
 \definecolor{YELLOW}{cmyk}{0,0,1,0}
 }

\makeatother

\begin{document}

\title{Shadowing in Compton scattering on nuclei}

\author{B.Z.~Kopeliovich}

\email{Boris.Kopeliovich@usm.cl}

\author{Ivan~Schmidt}

\email{Ivan.Schmidt@usm.cl}

\author{M.~Siddikov}

\email{Marat.Siddikov@usm.cl}

\affiliation{Departamento de F\'{\i}sica, Centro de Estudios Subat\'omicos, y Centro Cient\'ifico - Tecnol\'ogico de Valpara\'iso, Universidad T\'ecnica Federico Santa Mar\'{\i}a, Casilla 110-V, Valpara\'iso, Chile}

\preprint{USM-TH-260}
\begin{abstract}
We evaluate the shadowing effect in deeply virtual and real Compton
scattering on nuclei in the framework of the color dipole model. We
rely on the soft photon wave function derived in the instanton vacuum
model, and employ the impact parameter dependent phenomenological
elastic dipole amplitude. Both the effects of quark and the gluon
shadowing are taken into account. 
\end{abstract}
\maketitle

\section{Introduction}

Compton scattering, $\gamma^{*}+p\to\gamma+p$, with initial photons
both real or virtual, has been a subject of intensive theoretical
and experimental investigation~\cite{Mueller:1998fv,Ji:1996nm,Ji:1998pc,Radyushkin:1996nd,Radyushkin:1997ki,Radyushkin:2000uy,Ji:1998xh,Collins:1998be,Collins:1996fb,Brodsky:1994kf,Goeke:2001tz,Diehl:2000xz,Belitsky:2001ns,Diehl:2003ny,Belitsky:2005qn,Goncalves:2010ci}.
While in the case of deeply-virtual Compton scattering (DVCS), where
the initial photon is highly-virtual, the QCD factorization has been
proven \cite{Radyushkin:1997ki,Collins:1998be,Ji:1998xh} and the
amplitude can be expressed in terms of the generalized parton distributions
(GPD) \cite{Mueller:1998fv,Ji:1996nm,Ji:1998pc,Radyushkin:1996nd,Radyushkin:1997ki,Radyushkin:2000uy,Ji:1998xh,Collins:1998be,Collins:1996fb,Brodsky:1994kf,Goeke:2001tz,Diehl:2000xz,Belitsky:2001ns,Diehl:2003ny,Belitsky:2005qn},
in the case of real Compton scattering (RCS) the available theoretical
tools are rather undeveloped.

On the one hand, as it has been shown in~\cite{Kronfeld:1991kp,Brooks:2000nb},
for large momentum transfer $\Delta_{\perp}$ it is possible to factorize
the RCS amplitude~\cite{Lepage:1979zb,Lepage:1980fj} and express
it in terms of the distribution amplitudes of the proton. On the other
hand, it is possible to express the amplitude of the process via the
minus 1st-moment of GPDs at zero skewedness \cite{Radyushkin:1997ki,Diehl:1998kh,Diehl:2004cx}.

DVCS and RCS on a proton have been studied recently within the color
dipole approach in~\cite{Kopeliovich:2008ct,Kopeliovich:2009cx}.
Here we extend that study to nuclear targets. The DVCS process on
a nuclear target has been measured at HERA by HERMES collaboration~\cite{Airapetian:2009bi}
and may be also studied at the future Electron Ion Collider (EIC)
and Large Hadron Electron Collider (LHeC)~\cite{Klein:2008zza,Zimmermann:2008zzd}.
The Real Compton Scattering (RCS) may be measured at the LHC, as a
subprocess in hadron-hadron collisions in ultraperipheral kinematics.
Since both DVCS and RCS are studied in the high-energy kinematics,
the nuclear effects reveal themselves as shadowing corrections.

The general framework for evaluation of the shadowing corrections
is the Gribov-Glauber approach~\cite{Gribov:1968jf}. While in the
asymptotic high-energy ({}``frozen'') regime the shadowing corrections
were studied in~\cite{Machado:2008tp,Machado:2009cd}, we use an
approach which is also valid for  intermediate energies. Also, we
take into account the gluon shadowing corrections, which appear for
$x_{B}\lesssim10^{-3}$ and give a sizeable contribution for $x_{B}\sim10^{-5}$.

The paper is organized as follows. In Sections~\ref{sec:DipoleModel}
we review the general formalism of the color dipole approach. In Section~\ref{sec:frozen}
we discuss the frequently used frozen approximation which is valid
for asymptotically large energies. In Section~\ref{sec:sha-coh}
we discuss the method which will be used for calculations of nuclear
shadowing effects and demonstrate that for asymptotically large energies
it reproduces the results from~Section~\ref{sec:frozen}. In Section~\ref{sec:GluonShadowing}
we discuss the gluon shadowing and its effect on the DVCS and RCS
observables. In Section~\ref{sec:WFfromIVM} the wavefunction of
a real photon is evaluated in the instanton vacuum model. In Section~\ref{sec:Results}
we present the results of numerical evaluation, and in Section~\ref{sec:Conclusions}
we draw conclusions.

\section{Color dipole model}

\label{sec:DipoleModel} The color dipole model is particularly efficient
at high energies, where the dominant contribution to the Compton amplitude
comes from gluonic exchanges. Then the general expression for the
Compton amplitude on a nucleon has the form,

\begin{equation}
\mathcal{A}_{\mu\nu}^{(ij)}\left(s,\Delta\right)=e_{\mu}^{(i)}e_{\nu}^{(j)}\int\limits _{0}^{1}d\beta_{1}d\beta_{2}d^{2}r_{1}d^{2}r_{2}\bar{\Psi}_{f}^{(i)}\left(\beta_{2},\vec{r}_{2}\right)\mathcal{A}^{d}\left(\beta_{1},\vec{r}_{1};\beta_{2},\vec{r}_{2};\Delta\right)\Psi_{in}^{(j)}\left(\beta_{1},\vec{r}_{1}\right),\label{eq:Convolution:Full}\end{equation}
 where $e_{\mu}^{(i)}$ is the photon polarization vector; $\beta_{1,2}$
are the light-cone fractional momenta of the quark and antiquark,
$\vec{r}_{1,2}$ are the transverse distances in the final and initial
dipoles respectively; $\Delta$ is the momentum transfer in the Compton
scattering\textbf{,} $\mathcal{A}^{d}(...)$ is the scattering amplitude
of the dipole on the target (proton or nucleus), and $\Psi_{in(f)}^{(i)}\left(\beta,\vec{r}\right)$
are the wavefunctions of the initial and final photons in the polarization
state $i$ \cite{Dorokhov:2006qm}.

At high energies in the small angle approximation, $\Delta/\sqrt{s}\ll1$,
the quark separation and fractional momenta $\beta$ are preserved,
so, \begin{eqnarray}
\mathcal{A}^{d}\left(\beta_{1},\vec{r}_{1};\beta_{2},\vec{r}_{2};Q^{2},\Delta\right) & \approx & \delta\left(\beta_{1}-\beta_{2}\right)\delta\left(\vec{r}_{1}-\vec{r}_{2}\right)\int d^{2}b'\, e^{i\vec{\Delta}\vec{b}'}\Im m\, f_{\bar{q}q}^{N}(\vec{r}_{1},\vec{b}',\beta_{1})\label{eq:DVCSIm-BK}\\
\Im m\, f_{\bar{q}q}^{N}(\vec{r},\vec{b}',\beta) & = & \frac{1}{12\pi}\int\frac{d^{2}k\, d^{2}\Delta}{\left(k+\frac{\Delta}{2}\right)^{2}\left(k-\frac{\Delta}{2}\right)^{2}}\alpha_{s}\mathcal{F}\left(x,\vec{k},\vec{\Delta}\right)\, e^{i\vec{b}'\cdot\vec{\Delta}}\nonumber \\
 & \times & \left(e^{-i\beta\vec{r}\cdot\left(\vec{k}-\frac{\vec{\Delta}}{2}\right)}-e^{i(1-\beta)\vec{r}\cdot\left(\vec{k}-\frac{\vec{\Delta}}{2}\right)}\right)\left(e^{i\beta\vec{r}\cdot\left(\vec{k}+\frac{\vec{\Delta}}{2}\right)}-e^{-i(1-\beta)\vec{r}\cdot\left(\vec{k}+\frac{\vec{\Delta}}{2}\right)}\right),\label{eq:ImFN_GPD}\end{eqnarray}
where \begin{eqnarray}
\frac{\mathcal{F}\left(x,\vec{k},\vec{\Delta}\right)}{k^{2}} & \equiv & H_{g}\left(x,\vec{k},\vec{\Delta}\right)=\frac{1}{2}\int d^{2}r\, e^{ik\cdot r}\int\frac{dz^{-}}{2\pi}e^{ix\bar{P}^{+}z^{-}}\times\nonumber \\
 & \times & \left\langle P'\left|G_{+\alpha}\left(-\frac{z}{2},\,-\frac{\vec{r}}{2}\right)\gamma_{+}\mathcal{L}\left(-\frac{z}{2}-\frac{\vec{r}}{2},\frac{z}{2}+\frac{\vec{r}}{2}\right)G^{+\alpha}\left(\frac{z}{2},\,\frac{\vec{r}}{2}\right)\right|P\right\rangle \label{eq:UninteggratedGPD-definition}\end{eqnarray}
is the gluon GPD of the target, $P'=P+\Delta,$$\bar{P}=(P+P')/2$,
$G_{\mu\nu}(x)$ is the gluon loop operator, $\mathcal{L}_{\infty}\left(x,y\right)$
is the Wilson factor required by gauge covariance. For this GPD we
use a gaussian parameterization~\cite{Kopeliovich:2007fv,Kopeliovich:2008nx,Kopeliovich:2008da},

\begin{equation}
\mathcal{F}\left(x,\vec{k},\vec{\Delta}\right)=\frac{3\sigma_{0}(x)}{16\pi^{2}\alpha_{s}}\left(k+\frac{\Delta}{2}\right)^{2}\left(k-\frac{\Delta}{2}\right)^{2}R_{0}^{2}(x)\exp\left(-\frac{R_{0}^{2}(x)}{4}\left(\vec{k}^{2}+\frac{\vec{\Delta}^{2}}{4}\right)\right)\exp\left(-\frac{1}{2}B(x)\Delta^{2}\right),\label{eq:GPDParametrization}\end{equation}

where $\sigma_{0}(x),\, R_{0}^{2}(x),\, B(x)$ are the free parameters
fixed from the DIS and $\pi p$ scattering data. We shall discuss
them in more detail in Section~\ref{sec:Results}. The parameterization~(\ref{eq:GPDParametrization})
does not depend on the longitudinal momentum transfer and decreases
exponentially as a function of $\Delta^{2}$. Since the parameterization~(\ref{eq:GPDParametrization})
is an effective one and is valid only in the small-$x$ region, we
do not assume that it satisfies general requirements, such as positivity~\cite{Pobylitsa:2002gw}
and polynomiality~\cite{Ji:1998pc} constraints.

The prefactor $\left(k+\frac{\Delta}{2}\right)^{2}\left(k-\frac{\Delta}{2}\right)^{2}$
in~(\ref{eq:GPDParametrization}) guarantees convergence of the integrals
in the parameterization~(\ref{eq:DVCSIm-BK}). In the forward limit,
the amplitude (\ref{eq:DVCSIm-BK}) reduces to the saturated parameterization
of the dipole amplitude proposed by Golec-Biernat and W\"usthoff (GBW)~\cite{GolecBiernat:1998js},
\begin{eqnarray}
\sigma_{d}(\beta,r) & = & 2\int d^{2}b'\,\Im m\, f_{\bar{q}q}^{N}(\vec{r},\vec{b}',\beta)=\frac{1}{6\pi}\int\frac{d^{2}k}{k^{4}}\alpha_{s}\left(k^{2}\right)\mathcal{F}\left(x,\vec{k},\vec{0}\right)=\frac{\sigma_{0}(x)}{2}\left(1-\exp\left(-\frac{r^{2}}{R_{0}(x)}\right)\right)\label{eq:Sigma_QQ_GPD}\end{eqnarray}
 Generally, the amplitude $f_{\bar{q}q}^{N}(...)$ involves  nonperturbative
physics, but the asymptotic behaviour for small~$r$ is controlled
by pQCD \cite{Kopeliovich:1981}:\[
f_{\bar{q}q}^{N}(\vec{r},\vec{\Delta},\beta)_{r\to0}\propto r^{2},\]
 up to slowly varying corrections $\sim\ln(r)$.

Calculation of the differential cross section also involves the real
part of scattering amplitude, whose relation to the imaginary part
is quite straightforward. According to \cite{Bronzan:1974jh}, if
the limit $\lim\limits _{s\to\infty}\left(\frac{\mathcal{I}m\,{f}}{s^{\alpha}}\right)$
exists and is finite, then the real and imaginary parts of the forward
amplitude are related as\begin{equation}
\mathcal{R}e\,{f(\Delta=0)}=s^{\alpha}\tan\left[\frac{\pi}{2}\left(\alpha-1+\frac{\partial}{\partial\ln s}\right)\right]\frac{\Im m\,{f(\Delta=0)}}{s^{\alpha}}.\label{eq:BronzanFul}\end{equation}
 In the model under consideration the imaginary part of the forward
dipole amplitude indeed has a power dependence on energy, $\mathcal{I}m\, f(\Delta=0;\, s)\sim s^{\alpha}$,
so (\ref{eq:BronzanFul}) simplifies to \begin{eqnarray}
\frac{\mathcal{R}e\,\mathcal{A}}{\Im m\,\mathcal{A}} & =\tan\left(\frac{\pi}{2}(\alpha-1)\right)\equiv\epsilon.\end{eqnarray}

This fixes the phase of the forward Compton amplitude, which we retain
for nonzero momentum transfers, assuming similar dependences for the
real and imaginary parts. Finally we arrive at, \begin{equation}
\mathcal{A}_{\mu\nu}^{(ij)}=(\epsilon+i)e_{\mu}^{(i)}(q')e_{\nu}^{(j)}(q)\int d^{2}r\int\limits _{0}^{1}d\beta\,\bar{\Psi}_{f}^{(i)}(\beta,r)\Psi_{in}^{(j)}(\beta,r)\,\Im m\, f_{\bar{q}q}^{N}(\vec{r},\vec{\Delta},\beta,s),\label{eq:DVCS-Im-conv}\end{equation}

For the cross-section of unpolarized Compton scattering, from~(\ref{eq:DVCS-Im-conv})
we obtain, \begin{eqnarray}
\frac{d\sigma_{el}^{\gamma p}}{dt} & = & \frac{1+\epsilon^{2}}{16\pi}\sum_{ij}\left|\mathcal{A}_{\mu\nu}^{(ij)}\right|^{2}=\nonumber \\
 & = & \frac{1+\epsilon^{2}}{16\pi}\sum_{ij}\left|\int d^{2}r\int\limits _{0}^{1}d\beta\,\bar{\Psi}_{f}^{(i)}(\beta,r)\Psi_{in}^{(j)}(\beta,r)\,\Im m\, f_{\bar{q}q}^{N}(\vec{r},\vec{\Delta},\beta)\right|^{2}.\label{eq:DVCS-cross-section}\end{eqnarray}

\section{Nuclear shadowing in the "frozen" limit}

\label{sec:frozen}

Nuclear shadowing signals the closeness of the unitarity limit. Hard
reactions possess this feature only if they have a contribution from
soft interactions. In DIS and DVCS the soft contribution arises from
the so called aligned jet configurations \cite{bjorken}, corresponding
to  $\bar{q}q$ fluctuations very asymmetric in sharing the photon
momentum, $\beta\ll1$. Such virtual photon fluctuations, having large
transverse separation, are the source of shadowing \cite{k-povh}
.

Calculation of nuclear shadowing simplifies considerably in the case
of long coherence length \cite{krt2}, i.e. long lifetime of the photon
fluctuations, when it considerably exceeds the nuclear size. In this
case Lorentz time dilation "freezes" the
transverse size of the fluctuation during propagation though the nucleus.
Then the Compton amplitude of coherent scattering, which leaves the
nucleus intact, has the same form as Eq.~(\ref{eq:DVCS-Im-conv})
with a replacement of the nucleon Compton amplitude by the nuclear
one, \begin{equation}
\Im mf_{\bar{q}q}^{N}(r,\beta,\Delta)\Rightarrow{\cal \Im}mf_{\bar{q}q}^{A}(r,\beta,\Delta)=\int d^{2}b\, e^{i\vec{\Delta}\cdot\vec{b}}\,\left[1-e^{-{\cal \Im}mf_{\bar{q}q}^{N}(r,\beta,\Delta=0)\, T_{A}(b)}\right],\label{nucl-coh}\end{equation}
 where $b$ is impact parameter of the photon-nucleus collision, $T_{A}(b)=\int_{-\infty}^{\infty}dz\,\rho_{A}(b,z)$
is the nuclear thickness function, given by the integral of nuclear
density along the direction of the collisions. In this expression
we neglect the real part of the amplitude which is particularly small
for a coherent nuclear interaction.

For incoherent Compton scattering, which results in nuclear fragmentation
without particle production (quasielastic scattering), the cross section
has the form \cite{kps1},

\begin{eqnarray}
\frac{d\sigma_{qel}^{\gamma A}}{dt} & = & B_{el}\, e^{B_{el}t}\,\sum_{ij}\int\limits _{0}^{1}d\beta\int d^{2}rd^{2}r'\,\bar{\Psi}_{f}^{(i)}(\beta,r)\,\bar{\Psi}_{f}^{(i)}(\beta,r')\,\Psi_{in}^{(j)}(\beta,r)\,\Psi_{in}^{(j)}(\beta,r')\nonumber \\
 & \times & \exp\left[\frac{1}{2}\Bigl(\sigma_{\bar{q}q}(r)-\sigma_{\bar{q}q}(r')\Bigr)\, T_{A}(b)\right]\,\left\{ \exp\left[\frac{\sigma_{\bar{q}q}(r)\,\sigma_{\bar{q}q}(r')}{16\pi\, B_{el}}\, T_{A}(b)\right]-1\right\} \nonumber \\
 & \approx & \frac{e^{B_{el}t}}{16\pi}\int d^{2}b\, T_{A}\left(b\right)\left|\int\limits _{0}^{1}d\beta\, d^{2}r\,\bar{\Psi}_{f}\left(\beta,r\right)\sigma_{\bar{q}q}\left(r,s\right)\Psi_{in}\left(\beta,r\right)\exp\left[-\frac{1}{2}\sigma_{\bar{q}q}\left(r\right)T_{A}\left(b\right]\right)\right|^{2}.\label{eq:quasielastic}\end{eqnarray}
 Here $B_{el}$ is the $t$-slope of elastic dipole-nucleon amplitude.
In this equation we treated the term quadratic in the dipole cross
section as a small number and expanded the exponential in curly brackets.

\section{Onset of nuclear shadowing}

\label{sec:sha-coh}

\subsection{Coherent Compton scattering}

The regime of frozen dipole size discussed in the previous section
is valid only at very small $x_{B}$ in DVCS, or at high energies
in RCS. However, at medium small $x_{B}$ a dipole can "breath",
i.e. vary its size, during propagation through the nucleus, and one
should rely on a more sophisticated approach.

In this paper we employ the description of the onset of shadowing
developed in~\cite{Kopeliovich:1998gv} and based on the light-cone
Green function technique \cite{kz91}. The propagation of a color
dipole in a nuclear medium is described as  motion in an absorptive
potential, \emph{i.e.} \begin{equation}
i\,\frac{\partial W\left(z_{2},r_{2};z_{1},r_{1}\right)}{\partial z_{2}}=-\frac{\Delta_{r_{2}}W\left(z_{2},r_{2};z_{1},r_{1}\right)}{\nu\alpha(1-\alpha)}-\frac{i\rho_{A}\left(z_{2},r_{2}\right)\sigma_{\bar{q}q}\left(r_{2}\right)}{2}\, W\left(z_{2},r_{2};z_{1},r_{1}\right),\label{eq:W-definition}\end{equation}
 where the Green function $W\left(z_{2},r_{2};z_{1},r_{1}\right)$
describes the probability amplitude for propagation of dipole state
with size $r_{1}$ at the light-cone starting point $z_{1}$ to the
dipole state with size $r_{2}$ at the light-cone point $z_{2}.$
Then the shadowing correction to the amplitude has the form\begin{eqnarray*}
\delta A(s,\vec{\Delta}_{\perp}) & = & \int d^{2}b\, e^{i\vec{\Delta}_{}\cdot\vec{b}_{\perp}}\int\limits _{z_{1}\le z_{2}}dz_{1}dz_{2}\,\rho_{A}(b,z_{1})\,\rho_{A}(b,z_{2})\int\limits _{0}^{1}d\alpha d^{2}r_{1}d^{2}r_{2}\times\\
 &  & \bar{\Psi}_{f}\left(\alpha,r_{2}\right)\sigma_{\bar{q}q}\left(r_{2}\right)\, W\left(z_{2},r_{2};z_{1},r_{1}\right)\sigma_{\bar{q}q}\left(r_{1}\right)\Psi_{in}\left(\alpha,r_{1}\right)e^{ik_{min}\left(z_{2}-z_{1}\right)},\end{eqnarray*}
 where\[
k_{min}=\frac{Q^{2}\alpha(1-\alpha)+m_{q}^{2}}{2\nu\alpha(1-\alpha)}.\]

Equation~(\ref{eq:W-definition}) is quite complicated and in the
general case may be solved only numerically \cite{nemchik}. However
in some cases an analytic solution is possible. For example, in the
limit of long coherence length, $l_{c}\gg R_{A}$, relevant for high-energy
accelerators like LHC, one can neglect the {}``kinetic'' term~$\propto\Delta_{r_{2}}W\left(z_{2},r_{2};z_{1},r_{1}\right)$
in ~(\ref{eq:W-definition}), and get the Green function in the "frozen"
approximation \cite{kz91}, \begin{equation}
W\left(z_{2},r_{2};z_{1},r_{1}\right)=\delta^{2}\left(r_{2}-r_{1}\right)\exp\left(-\frac{1}{2}\sigma_{\bar{q}q}\left(r_{1}\right)\int\limits _{z_{1}}^{z_{2}}d\zeta\,\rho_{A}\left(\zeta,b\right)\right).\label{eq:W-frozen}\end{equation}
 Then the shadowing correction~(\ref{eq:DVCS-COH-SHA}) simplifies
to \begin{eqnarray}
\delta\mathcal{A}(s,\Delta_{\perp}) & = & \int d^{2}b\, e^{i\vec{\Delta}_{\perp}\cdot\vec{b}_{\perp}}\int\limits _{z_{1}\le z_{2}}dz_{1}dz_{2}\,\rho_{A}\left(z_{1},b\right)\rho_{A}\left(z_{2},b\right)\int\limits _{0}^{1}d\alpha\, d^{2}r\,\sigma_{\bar{q}q}^{2}\left(r,b\right)\times\label{eq:DVCS-COH-SHA}\\
 &  & \bar{\Psi}_{f}\left(\alpha,r\right)\exp\left(-\frac{1}{2}\sigma_{\bar{q}q}\left(r\right)\int\limits _{z_{1}}^{z_{2}}d\zeta\rho_{A}\left(\zeta,b\right)\right)\Psi_{in}\left(\alpha,r\right)e^{ik_{min}\left(z_{2}-z_{1}\right)}.\nonumber \end{eqnarray}
 If we neglect the real part of the amplitude and the longitudinal
momentum transfer $k_{min}$ (which is justified for asymptotically
large $s$), and average over polarizations, then taking the integral
over $z_{1,2}$ ''by parts'' in~(\ref{eq:DVCS-COH-SHA}), we get
for the elastic amplitude \begin{eqnarray}
\mathcal{A}(s,\Delta_{\perp}) & = & 2\int d^{2}b\, e^{i\vec{\Delta}_{\perp}\cdot\vec{b}_{\perp}}\int\limits _{0}^{1}d\alpha\, d^{2}r\,\bar{\Psi}_{f}\left(\alpha,r\right)\left[1-\exp\left(-\frac{1}{2}\sigma_{\bar{q}q}\left(r\right)\int\limits _{-\infty}^{+\infty}d\zeta\rho_{A}\left(\zeta,b\right)\right)\right]\Psi_{in}\left(\alpha,r\right).\label{eq:A-frozen}\end{eqnarray}
Another case where an analytical solution is possible is when the
effective dipole sizes are small and the function $\sigma_{\bar{q}q}(r)$
may be approximated as \begin{equation}
\sigma_{\bar{q}q}(r)\approx Cr^{2}.\label{eq:sigma-small-r}\end{equation}
 This approximation cannot be precise even at high virtuality $Q^{2}$
in DVCS, since there are contributions of the aligned jet configurations
mentioned above, which permit large dipoles even for large $Q^{2}$.
Moreover, such aligned jet configurations of the dipole provide the
main contribution to nuclear shadowing \cite{k-povh}. Nevertheless,
for the sake of simplicity we rely on this approximation in order
to estimate the magnitude of the shadowing corrections. Then Eq.~(\ref{eq:W-definition})
yields for $W\left(z_{2},r_{2};z_{1},r_{1}\right)$ the well-known
evolution operator of harmonic oscillator, although with complex frequency\\
 \begin{eqnarray}
W\left(z_{2},r_{2};z_{1},r_{1}\right) & = & \frac{a}{2\pi i\sin\left(\omega\Delta z\right)}\exp\left(\frac{ia}{2\sin\left(\omega\Delta z\right)}\left[\left(r_{1}^{2}+r_{2}^{2}\right)\cos\left(\omega\Delta z\right)-2\vec{r}_{1}\cdot\vec{r}_{2}\right]\right),\label{eq:W-oscillator}\\
\omega^{2} & = & \frac{-2iC\rho_{A}}{\nu\alpha(1-\alpha)},\nonumber \\
a^{2} & = & -iC\rho_{A}\nu\alpha(1-\alpha)/2\nonumber \end{eqnarray}

Notice that for DVCS in the kinematics of the HERMES experiment~\cite{Airapetian:2009bi}
the coherence length \[
l_{c}\approx\frac{1}{2m_{N}x_{B}}\sim1.7\, fm\]
 is comparable with the mean inter-nucleon spacing in nuclei and is
much smaller than the radii of heavy nuclei. Therefore the frozen
approximation employed in \cite{Machado:2008tp} cannot be used for
an interpretation of HERMES data, instead one should rely on the Green
function method described above.

\subsection{Incoherent scattering}

\label{sec:sha-incoh}

In addition to the coherent processes which leave the recoil nucleus
intact, a large contribution to photon production comes from incoherent
Compton scattering, where the target nucleus breaks up. Using the
missing mass technique one can select events $\gamma A\to\gamma A^{*}$
in which the nucleus breaks up to fragments without production of
mesons. In this case one can employ completeness of the final states,
which greatly simplifies the calculations.

The analysis of such processes in electroproduction of vector mesons
was done in~\cite{Kopeliovich:2001xj}, and its extension to the
DVCS and RCS is quite straightforward and yields

\begin{eqnarray}
\left.\frac{d\sigma}{dt}\right|_{\Delta_{\perp}=0} & = & \frac{1}{16\pi}\int d^{2}bdz\,\rho_{A}\left(b,z\right)\left|F_{1}\left(b,z\right)-F_{2}\left(b,z\right)\right|^{2},\label{xsection}\\
F_{1}\left(b,z\right) & = & \int\limits _{0}^{1}d\alpha\, d^{2}r_{1}\, d^{2}r_{2}\,\bar{\Psi}_{f}\left(\alpha,\vec{r}_{2}\right)W\left(+\infty,\vec{r}_{2};z,\vec{r}_{1}\right)\sigma_{\bar{q}q}\left(\vec{r}_{1},s\right)\Psi_{in}\left(\alpha,\vec{r}_{1}\right),\\
F_{2}\left(b,z\right) & = & \int\limits _{-\infty}^{z}dz_{2}\int\limits _{0}^{1}d\alpha\, d^{2}r_{1}\, d^{2}r_{2}\, d^{2}r_{3}\,\bar{\Psi}_{f}\left(\alpha,\vec{r}_{3}\right)W\left(+\infty,\vec{r}_{3};z,\vec{r}_{2}\right)\sigma_{\bar{q}q}\left(\vec{r}_{2},s\right)\times\label{eq:incoh-frozen}\\
 &  & W\left(z,\vec{r}_{2};z_{2},\vec{r}_{1}\right)\rho_{A}\left(z_{2},b\right)\sigma_{\bar{q}q}\left(\vec{r}_{1},s\right)\Psi_{in}\left(\alpha,\vec{r}_{1}\right).\nonumber \end{eqnarray}
 At sufficiently high energies one can rely on the frozen approximation
introduced in the previous section, and this formula may be simplified
\cite{hkz}, \begin{eqnarray}
F_{1}\left(b,z\right) & = & \int\limits _{0}^{1}d\alpha d^{2}r\bar{\Psi}_{f}\left(\alpha,r\right)\sigma_{\bar{q}q}\left(r_{1},s\right)\Psi_{in}\left(\alpha,r\right)\exp\left(-\frac{1}{2}\sigma_{\bar{q}q}\left(r\right)\int\limits _{z}^{+\infty}d\zeta\rho_{A}\left(\zeta,b\right)\right),\\
F_{2}\left(b,z\right) & = & \int\limits _{0}^{1}d\alpha d^{2}r\bar{\Psi}_{f}\left(\alpha,r\right)\times\\
 &  & \left[\exp\left(-\frac{1}{2}\sigma_{\bar{q}q}\left(r\right)\int\limits _{z}^{+\infty}d\zeta\rho_{A}\left(\zeta,b\right)\right)-\exp\left(-\frac{1}{2}\sigma_{\bar{q}q}\left(r\right)\int\limits _{-\infty}^{+\infty}d\zeta\rho_{A}\left(\zeta,b\right)\right)\right]\sigma_{\bar{q}q}\left(r,s\right)\Psi_{in}\left(\alpha,r\right),\nonumber \end{eqnarray}

Correspondingly the cross section (\ref{xsection}) takes the form,
\begin{eqnarray}
\left.\frac{d\sigma}{dt}\right|_{\Delta_{\perp}=0} & = & \frac{1}{16\pi}\int d^{2}b\, T_{A}\left(b\right)\left|\int\limits _{0}^{1}d\alpha d^{2}r\bar{\Psi}_{f}\left(\alpha,r\right)\sigma_{\bar{q}q}\left(r,s\right)\Psi_{in}\left(\alpha,r\right)\exp\left(-\frac{1}{2}\sigma_{\bar{q}q}\left(r\right)T_{A}\left(b\right)\right)\right|^{2}.\label{qel}\end{eqnarray}
 This expression reproduces Eq.~(\ref{eq:quasielastic}) derived
in the "frozen" limit. It is easy to see
that in the limit of a transparent nucleus, $\sigma_{\bar{q}q}T_{A}\ll1$,
the cross section Eq.~(\ref{qel}) rises linearly  with $A$. However
in the limit of a very opaque nucleus (black disk limit), $\sigma_{\bar{q}q}T_{A}\gg1$,
the absorptive exponential factor in (\ref{qel}) terminates the contribution
of central impact parameters, and $d\sigma/dt|_{\Delta_{\perp}=0}\propto A^{1/3}$.

In case of the approximation~(\ref{eq:sigma-small-r}), we may use
the explicit expression~(\ref{eq:W-oscillator}) for the Green function
$W\left(z_{2},\vec{r}_{2};z_{1},\vec{r}_{1}\right)$ to get

\begin{eqnarray}
F_{1}\left(b,z\right) & = & \int\limits _{0}^{1}d\alpha d^{2}r_{1}d^{2}r_{2}\bar{\Psi}_{f}\left(\alpha,\vec{r}_{2}\right)\frac{a}{2\pi i\sin\left(\omega\Delta z\right)}\exp\left(\frac{ia}{2\sin\left(\omega\Delta z\right)}\left[\left(r_{1}^{2}+r_{2}^{2}\right)\cos\left(\omega\Delta z\right)-2\vec{r}_{1}\cdot\vec{r}_{2}\right]\right)_{\Delta z=z_{\infty}-z}\\
 & \times & \sigma_{\bar{q}q}\left(\vec{r}_{1},s\right)\Psi_{in}\left(\alpha,\vec{r}_{1}\right),\nonumber \\
F_{2}\left(b,z\right) & = & \int\limits _{-\infty}^{z}dz_{2}\int\limits _{0}^{1}d\alpha\, d^{2}r_{1}d^{2}r_{2}d^{2}r_{3}\bar{\Psi}_{f}\left(\alpha,\vec{r}_{3}\right)\sigma_{\bar{q}q}\left(\vec{r}_{2},s\right)\rho_{A}\left(b,z_{2}\right)\sigma_{\bar{q}q}\left(\vec{r}_{1},s\right)\label{eq:F2-incoh-res}\\
 & \times & \frac{a}{2\pi i\sin\left(\omega\Delta z\right)}\exp\left(\frac{ia}{2\sin\left(\omega\Delta z\right)}\left[\left(r_{3}^{2}+r_{2}^{2}\right)\cos\left(\omega\Delta z\right)-2\vec{r}_{3}\cdot\vec{r}_{2}\right]\right)_{\Delta z=z_{\infty}-z_{2}}\nonumber \\
 & \times & \frac{a}{2\pi i\sin\left(\omega\Delta z\right)}\exp\left(\frac{ia}{2\sin\left(\omega\Delta z\right)}\left[\left(r_{1}^{2}+r_{2}^{2}\right)\cos\left(\omega\Delta z\right)-2\vec{r}_{1}\cdot\vec{r}_{2}\right]\right)_{\Delta z=z_{2}-z}\Psi_{in}\left(\alpha,\vec{r}_{1}\right).\nonumber \end{eqnarray}

\section{Gluon shadowing}

\label{sec:GluonShadowing}

It has been known since~\cite{Kancheli:1973vc} that in addition
to the quark shadowing inside nuclei there is also a shadowing of
gluons, which leads to attenuation of the gluon parton distributions.
While nuclear shadowing of quarks is directly measured in DIS, the
shadowing of gluons is poorly known from data~\cite{Arleo:2007js,Armesto:2006ph},
mainly due to the relatively large error bars in the nuclear structure
functions and their weak dependence on the gluon distributions, which
only comes via evolution. The theoretical predictions for the gluon
shadowing strongly depend on the implemented model--while for $x_{B}\gtrsim10^{-3}$
they all predict that the gluon shadowing is small, for $x_{B}\lesssim10^{-3}$
the predictions vary in a wide range (see the review~\cite{Armesto:2006ph}
and references therein). Since in this paper we also make predictions
for the LHC energy range, the gluon shadowing corrections should be
taken into account as well.

In the framework of the color dipole model the gluon attenuation factor
$R_{g}$ was evaluated in the Gribov-Glauber approach in~\cite{Kopeliovich:2001hf}.
It was found convenient to evaluate the gluon attenuation ratio $R_{g}$
defined as \[
R_{g}\left(x,Q^{2},b\right)=\frac{G_{A}\left(x,Q^{2},b\right)}{T_{A}(b)G_{N}\left(x,Q^{2},b\right)},\]
 where $G_{N}\left(x,Q^{2},b\right)$ is the impact parameter dependent
gluon GPD, relating it to the shadowing corrections in DIS with longitudinally
polarized photons, \begin{equation}
R_{g}\left(x,Q^{2},b\right)\approx1-\frac{\Delta\sigma_{L}^{\gamma^{*}p}\left(x,Q^{2},b\right)}{T_{A}(b)\sigma_{L}^{\gamma^{*}p}\left(x,Q^{2}\right)},\label{eq:R_g_damping}\end{equation}
 where $\Delta\sigma_{L}^{\gamma^{*}p}=\sigma_{L}^{\gamma^{*}A}-A\sigma_{L}^{\gamma^{*}p}$
is the shadowing correction at impact parameter $b$, and $\sigma_{L}^{\gamma^{*}p}\left(x,Q^{2}\right)$
is the total photoabsorption~cross section for a longitudinal photon.
The process with longitudinal photons is chosen because the aligned
jets configurations are suppressed by powers of $Q^{2}$, so that
the average size of the dipole is small, $\left\langle r^{2}\right\rangle \sim1/Q^{2},$
and nuclear shadowing mainly originates from gluons.

As it was shown in~\cite{Kopeliovich:2001hf},

\begin{equation}
\Delta\sigma_{L}^{\gamma^{*}p}\left(x,Q^{2},b\right)=\int_{-\infty}^{+\infty}dz_{1}\int_{-\infty}^{+\infty}dz_{2}\Theta\left(z_{2}-z_{1}\right)\rho_{A}\left(b,z_{1}\right)\rho_{A}\left(b,z_{2}\right)\Gamma\left(x,Q^{2},z_{2}-z_{1}\right),\label{eq:DeltaSigmaGluon}\end{equation}
 where $\rho_{A}(b,z)$ is the nuclear density, and~$\Gamma\left(x,Q^{2},\Delta z\right)$
is defined as \begin{eqnarray*}
\Gamma\left(x,Q^{2},\Delta z\right) & = & \Re e\int_{x}^{0.1}\frac{d\alpha_{G}}{\alpha_{G}}\frac{16\alpha_{em}\left(\sum Z_{q}^{2}\right)\alpha_{s}\left(Q^{2}\right)C_{eff}^{2}}{3\pi^{2}Q^{2}\tilde{b}^{2}}\times\\
 & \times & \left[\left(1-2\zeta-\zeta^{2}\right)e^{-\zeta}+\zeta^{2}(3+\zeta)E_{1}(\zeta)\right]\\
 & \times & \left[\frac{t}{w}+\frac{\sinh\left(\Omega\Delta z\right)}{t}\ln\left(1-\frac{t^{2}}{u^{2}}\right)+\frac{2t^{3}}{uw^{2}}+\frac{t\sinh\left(\Omega\Delta z\right)}{w^{2}}+\frac{4t^{3}}{w^{3}}\right],\end{eqnarray*}
 with \begin{eqnarray*}
\tilde{b}^{2} & = & \left(0.65GeV\right)^{2}+\alpha_{G}Q^{2},\\
\Omega & = & \frac{iB}{\alpha_{G}\left(1-\alpha_{G}\right)\nu},\\
B & = & \sqrt{\tilde{b}^{4}-i\alpha_{G}\left(1-\alpha_{G}\right)\nu C_{eff}\rho_{A}},\\
\nu & = & \frac{Q^{2}}{2m_{N}x},\\
\zeta & = & ixm_{N}\Delta z,\\
t & = & \frac{B}{\tilde{b}^{2}},\\
u & = & t\cosh\left(\Omega\Delta z\right)+\sinh\left(\Omega\Delta z\right),\\
w & = & \left(1+t^{2}\right)\sinh\left(\Omega\Delta z\right)+2t\cosh\left(\Omega\Delta z\right),\end{eqnarray*}
 For  heavy nuclei we may rely on the hard sphere approximation, $\rho_{A}(r)\approx\rho_{A}(0)\Theta\left(R_{A}-r\right)$,
and simplify~(\ref{eq:DeltaSigmaGluon}) to: \begin{eqnarray*}
\Delta\sigma^{\gamma^{*}p}\left(x,Q^{2},b\right) & \approx & \rho_{A}^{2}(0)\int_{0}^{L}d\Delta z(L-\Delta z)\Gamma\left(x,Q^{2},\Delta z\right),\end{eqnarray*}
 where $L=2\sqrt{R_{A}^{2}-b^{2}}.$ For the total cross-section after
integration over $\int d^{2}b$ we may get \begin{eqnarray*}
\Delta\sigma^{\gamma^{*}p}\left(x,Q^{2}\right) & = & \int d^{2}b\Delta\sigma^{\gamma^{*}p}\left(x,Q^{2},b\right)\\
 & \approx & \frac{\pi\rho_{A}^{2}(0)}{12}\int_{0}^{2R}d\Delta z\Gamma\left(x,Q^{2},\Delta z\right)\left(16R_{A}^{3}-12R_{A}^{2}\Delta z+\Delta z^{3}\right)\end{eqnarray*}
 The results of evaluation of the gluon shadowing are presented in
 Section~\ref{sec:Results}.

\section{Wavefunctions from the instanton vacuum}

\label{sec:WFfromIVM}In this section we present briefly some details
of evaluation of the wavefunction in the instanton vacuum model~(see~\cite{Schafer:1996wv,Diakonov:1985eg,Diakonov:1995qy}
and references therein). The central object of the model is the effective
action for the light quarks in the instanton vacuum, which in the
leading order in $N_{c}$ has the form~\cite{Diakonov:1985eg,Diakonov:1995qy}
\[
S=\int d^{4}x\left(\frac{N}{V}\ln\lambda+2\Phi^{2}(x)-\bar{\psi}\left(\hat{p}+\hat{v}-m-c\bar{L}f\otimes\Phi\cdot\Gamma_{m}\otimes fL\right)\psi\right),\]
 where $\Gamma_{m}$ is one of the matrices, $\Gamma_{m}=1,i\vec{\tau},\gamma_{5},i\vec{\tau}\gamma_{5}$,~$\psi$
and $\Phi$ are the fields of constituent quarks and mesons respectively,
$N/V$ is the density of the instanton gas, $\hat{v}\equiv v_{\mu}\gamma^{\mu}$
is the external vector current corresponding to the photon, $L$ is
the gauge factor, \begin{equation}
L\left(x,z\right)=P\exp\left(i\int\limits _{z}^{x}d\zeta^{\mu}v_{\mu}(\zeta)\right),\label{eq:L-factor}\end{equation}
 which provides the gauge covariance of the action, and $f(p)$ is
the Fourier transform of the zero-mode profile.

In the leading order in $N_{c}$, we have the same Feynman rules as
in  perturbative theory, but with a momentum-dependent quark mass
$\mu(p)$ in the quark propagator\begin{eqnarray}
S(p) & = & \frac{1}{\hat{p}-\mu(p)+i0}.\end{eqnarray}
 The mass of the constituent quark has a form \[
\mu(p)=m+M\, f^{2}(p),\]
 where $m\approx5$~MeV is the current quark mass, $M\approx350$~MeV
is the dynamical mass generated by the interaction with the instanton
vacuum background. Due to the presence of the instantons the coupling
of a vector current to a quark is also modified,\begin{eqnarray}
\hat{v} & \equiv & v_{\mu}\gamma^{\mu}\rightarrow\hat{V}=\hat{v}+\hat{V}^{nonl},\nonumber \\
\hat{V}^{nonl} & \approx & -2Mf(p)\frac{df(p)}{dp_{\mu}}v_{\mu}(q)+\mathcal{O}\left(q^{2}\right).\label{eq:V-nonl-expanded}\end{eqnarray}
 Notice that for an arbitrary photon momentum $q$ the expression
for $\hat{V}^{nonl}$ depends on the choice of the path in~(\ref{eq:L-factor})
and as a result one can find in the literature different expressions
used for evaluations~\cite{Dorokhov:2006qm,Anikin:2000rq,Dorokhov:2003kf,Goeke:2007j}.
In the limit $p\to\infty$ the function $f(p)$ falls off as $\sim\frac{1}{p^{3}},$
so for large $p\gg\rho^{-1}$, where $\rho\approx(600\, MeV)^{-1}$
is the mean instanton size, the mass of the quark $\mu(p)\approx m$
and vector current interaction vertex $\hat{V}\approx\hat{v}$. However
we would like to emphasize that the wavefunction $\Psi(\beta,r)$
gets contribution from both the soft and the hard parts, so even in
the large-$Q$ limit the instanton vacuum function is different from
the well-known perturbative result.

We have to evaluate the wavefunctions associated with the following
matrix elements:

\begin{eqnarray}
I_{\Gamma}(\beta,\vec{r}) & = & \int\frac{dz^{-}}{2\pi}e^{i\left(\beta+\frac{1}{2}\right)q^{-}z^{+}}\left\langle 0\left|\bar{\psi}\left(-\frac{z}{2}n-\frac{\vec{r}}{2}\right)\Gamma\psi\left(\frac{z}{2}n+\frac{\vec{r}}{2}\right)\right|\gamma(q)\right\rangle ,\end{eqnarray}
 where $\Gamma$ is one of the matrices $\Gamma=\gamma_{\mu},\gamma_{\mu}\gamma_{5},\sigma_{\mu\nu}.$
In the leading order in $N_{c}$ one can easily obtain\begin{equation}
I_{\Gamma}=\int\frac{d^{4}p}{(2\pi)^{4}}e^{i\vec{p}_{\perp}\vec{r}_{\perp}}\delta\left(p^{+}-\left(\beta+\frac{1}{2}\right)q^{+}\right)Tr\left(S(p)\hat{V}S(p+q)\Gamma\right).\label{eq:WF-LO}\end{equation}
 The evaluation of~(\ref{eq:WF-LO}) is quite tedious but straightforward.
Details of this evaluation may be found in~\cite{Dorokhov:2006qm}.

The overlap of the initial and final photon wavefunctions in~(\ref{eq:DVCS-cross-section})
was evaluated according to\begin{equation}
\Psi^{(i)*}(\beta,r,Q^{2}=0)\Psi^{(i)}(\beta,r,Q^{2})=\sum_{\Gamma}I_{\Gamma}^{*}\left(\beta,r^{*},0\right)I_{\Gamma}\left(\beta,r,Q^{2}\right),\label{eq:ConvolutionWF}\end{equation}
 where summation is done over possible polarization states $\Gamma=\gamma_{\mu},\gamma_{\mu}\gamma_{5},\sigma_{\mu\nu}$.
In the final state we should use $r_{\mu}^{*}=r_{\mu}+n_{\mu}\frac{q_{\perp}'\cdot r_{\perp}}{q_{+}}=r_{\mu}-n_{\mu}\frac{\Delta_{\perp}\cdot r_{\perp}}{q_{+}}$,
which is related to the reference frame with $q'_{\perp}=0$ in which
the components~(\ref{eq:WF-LO}) were evaluated.

\section{Numerical results}

\label{sec:Results}

In this section we present the results of numerical calculations.
 In this paper we consider two processes--DVCS and RCS. While physically
they differ only by the kinematics, the parameterizations used for
scattering with soft and hard photons  are different. For example,
for DVCS we have photons with large $Q^{2}$--in this region we have
Bjorken scaling, so all the model parameters such as basic cross-section
$\sigma_{0}$ and saturation radius $R_{0}$ in Eq.~(\ref{eq:GPDParametrization}).
should depend on the Bjorken $x_{B}.$ A widely accepted parameterization
which incorporates this features is the GBW parameterization~\cite{GolecBiernat:1998js}.
On the contrary, for RCS when $Q^{2}$ is vanishingly small, $x_{B}$
vanishes and thus all variables should depend on Mandelstam $s$.
An example of such parameterization is KST parameterization introduced
in \cite{kst2,Kopeliovich:2007fv,Kopeliovich:2008nx,Kopeliovich:2007wx}.
Up to the best of our knowledge there is no single parameterization
which incorporates both asymptotic cases. For this reason for DVCS
we used the GBW-style parameterization~\cite{GolecBiernat:1998js,Kopeliovich:2008da}
for the nonintegrated gluon density Eq.(\ref{eq:GPDParametrization}),
which has a form \begin{eqnarray}
\sigma_{0}(x) & = & 23.03\, mb=const,\\
R_{0}(x) & = & 0.4\, fm\times(x/x_{0})^{0.144},\label{eq:-2}\\
B(x) & = & B_{\gamma^{*}p\to\rho p}-\frac{1}{8}\, R_{0}^{2}(x),\label{eq:-1}\end{eqnarray}
where $x_{0}=3.04\times10^{-4}$, and $B_{\gamma^{*}p\to\rho p}(x,Q^{2}\gg1GeV^{2})\approx5\, GeV^{-2}$~\cite{zeus}.

\begin{figure}
\includegraphics[scale=0.4]{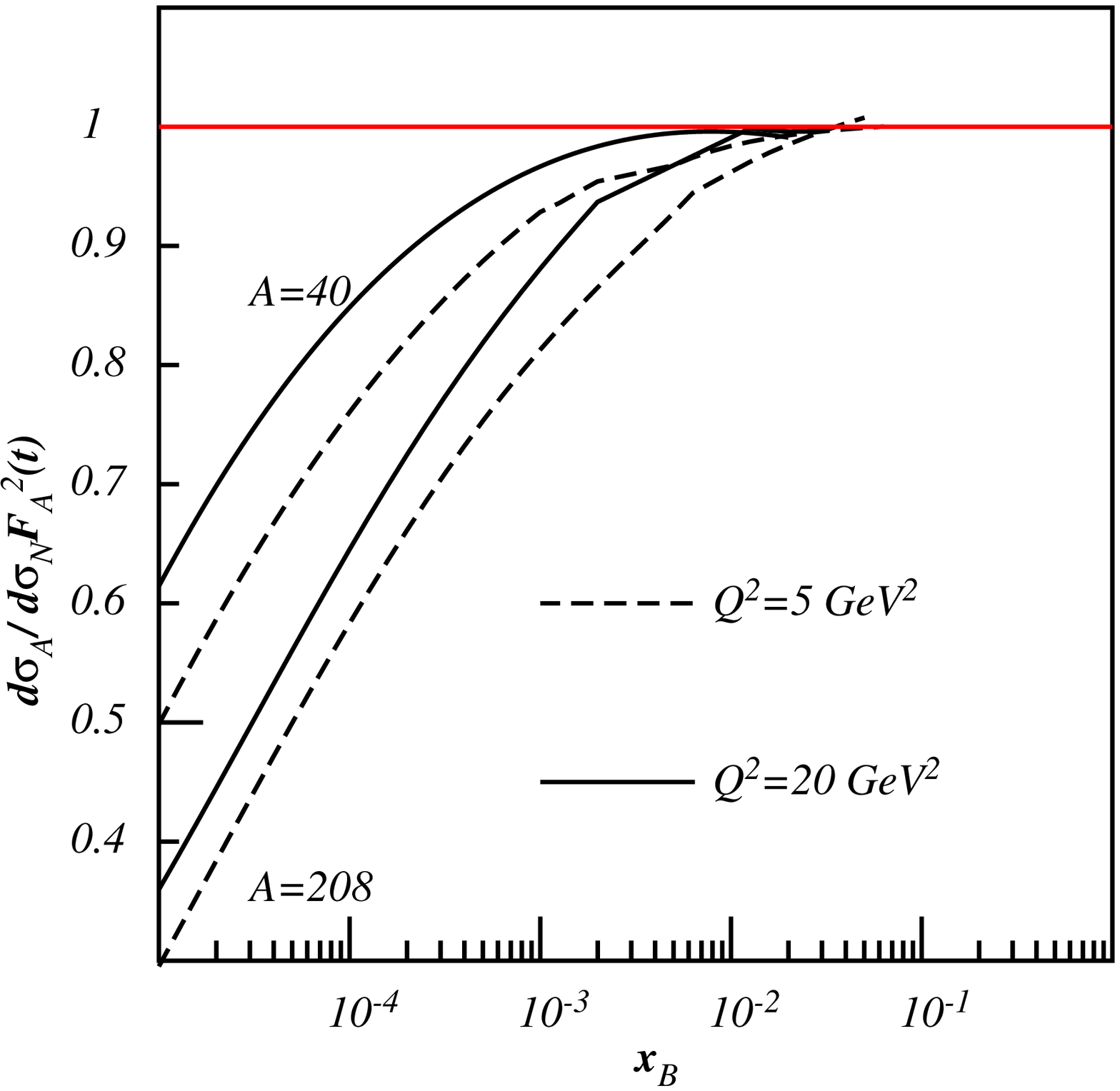}\includegraphics[scale=0.4]{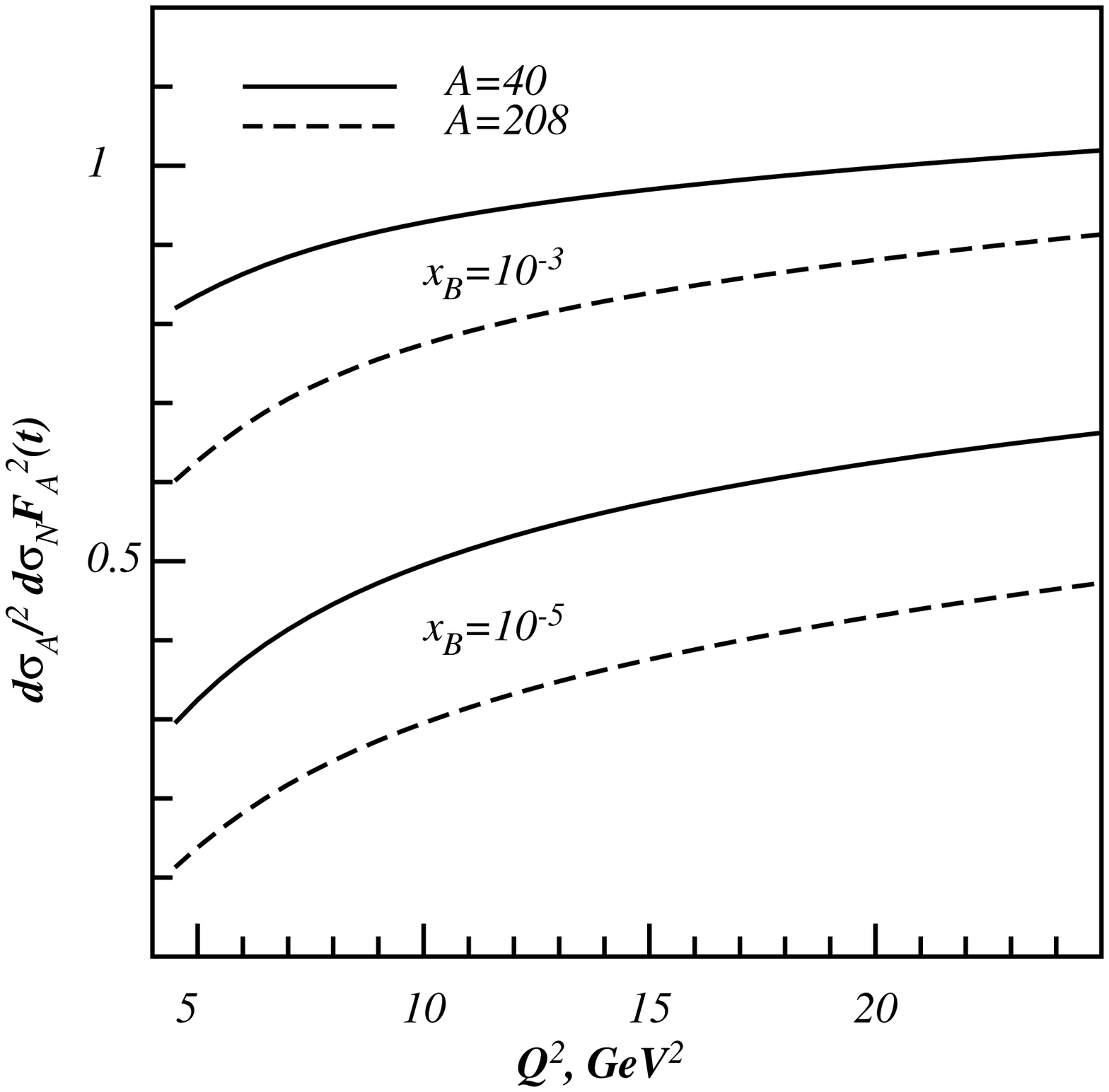}

\caption{\label{fig:DVCS-COH-RES} The nucleus to nucleon cross section ratio
for coherent DVCS as function of different kinematical variables.
Left: $x_{B}$-dependence of the shadowing, $t=t_{min}$, for different
$Q^{2}$ and $A$. From bottom to top: $Q^{2}=5$~GeV$^{2}$ and
$Q^{2}=20$~GeV$^{2}$. Right: $Q^{2}$-dependence, $t=-0.01$GeV$^{2}$,
for different $x_{B}$ and $A$. From bottom to top: $x_{B}=10^{-5}$
and $x_{B}=10^{-3}$.}

\end{figure}

For RCS we used a KST-style parameterization, which has the form ~\cite{kst2,Kopeliovich:2007fv,Kopeliovich:2008nx,Kopeliovich:2008ct,Kopeliovich:2009cx,Kopeliovich:2007wx,Kopeliovich:2008da}.\begin{eqnarray}
\sigma_{0}(s) & = & \sigma_{\pi p}(s)\left(1+\frac{3}{8}\frac{R_{0}^{2}(s)}{r_{\pi}^{2}}\right),\label{eq:dipole-2}\\
\sigma_{\pi p}(s) & = & \left(23.6\left(\frac{s}{s_{0}}\right)^{0.079}+1.45\left(\frac{s_{0}}{s}\right)^{0.45}\right)mb,\label{eq:dipole-3}\\
R_{0}(s) & = & 0.88\, fm\left(\frac{s_{0}}{s+s_{1}}\right)^{2},\label{eq:dipole-4}\\
B(s) & = & B_{\gamma^{*}p\to\rho p}-\frac{1}{8}\, R_{0}^{2}(s),\end{eqnarray}
where $s_{0}\approx1000\, GeV^{2},\: s_{1}\approx3600\, GeV^{2}$.
Note that in the large-$s$ region considered in this paper we may
neglect the second (Reggeon) term in Eq.~(\ref{eq:dipole-3}) and
set $s_{1}\approx0$ in~(\ref{eq:dipole-4}). As one can see from
the Figure~\ref{fig:DVCS-COH-RES}, the shadowing correction is increasing
towards small $x_{B}$, and for $x_{B}\sim10^{-5}$ the nuclear cross-section
ratio decreases by a factor of two compared to the naive estimate
$d\sigma_{A}\sim F_{A}^{2}(t)d\sigma_{N}$. As a function of the momentum
transfer $t$, the shadowing correction reveals the behavior qualitatively
similar to the nuclear formfactor $F_{A}(t)$: it steeply drops at
small-$t$ and has zeros for some $t$. Notice, however, that the
zero positions in the cross-section do not coincide with the zeros
of the formfactor. This is a result of shadowing which suppresses
the contribution of the central part of the nucleus and modifies the
$b$-dependence of the cross section compared to the formfactor. Notice
that, as was discussed in~\cite{Goeke:2009tu}, if we had sufficiently
high resolution in $t$, it would be possible to measure the contributions
of the {}``pure'' DVCS, without contributions from the Bethe-Heitler,
near the zeros of the formfactor.

The $Q^{2}$-dependence of the nuclear ratio is shown in the right
pane of the Figure~\ref{fig:DVCS-COH-RES}.

\begin{figure}[h]
\includegraphics[scale=0.4]{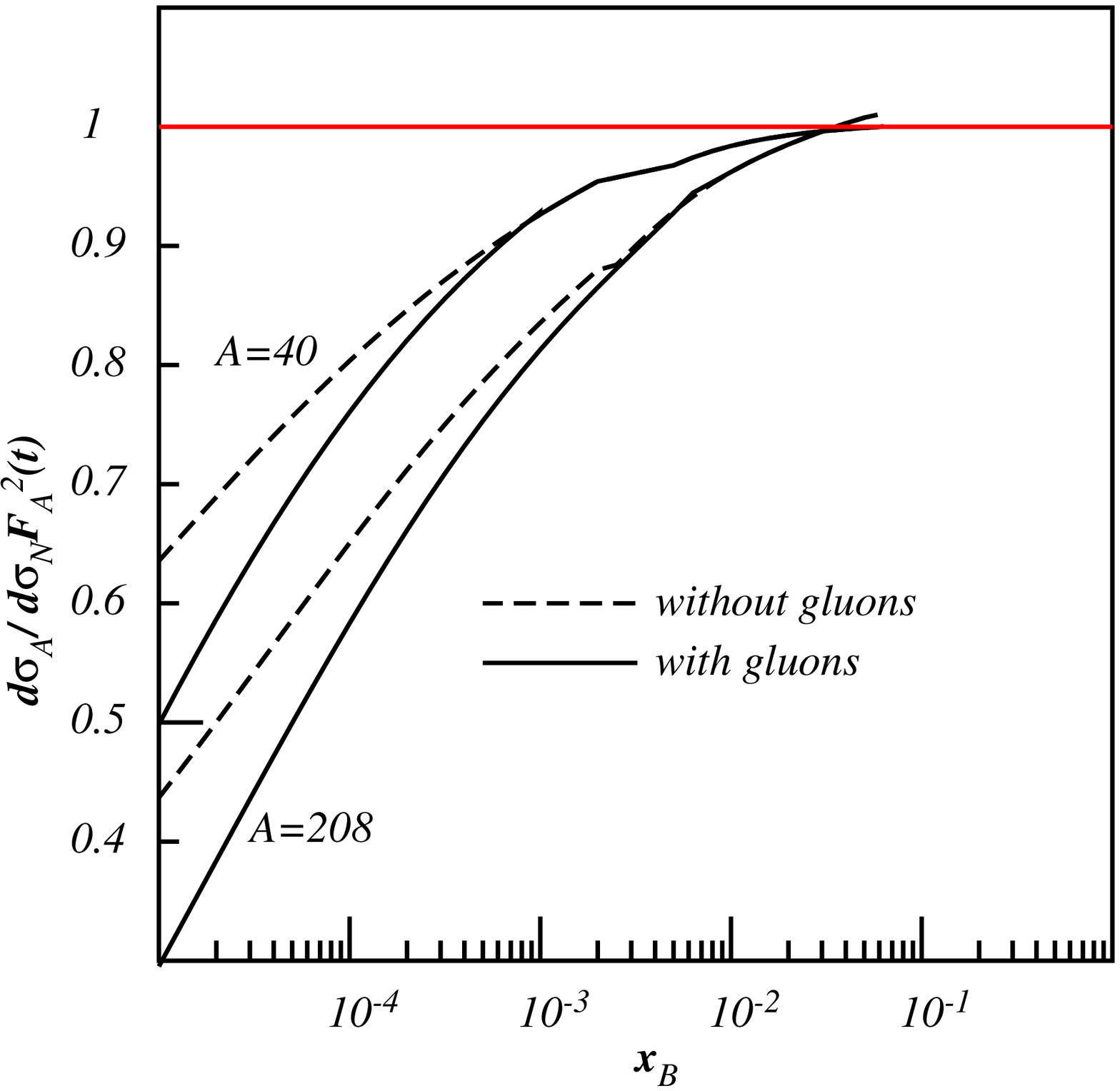}\includegraphics[scale=0.4]{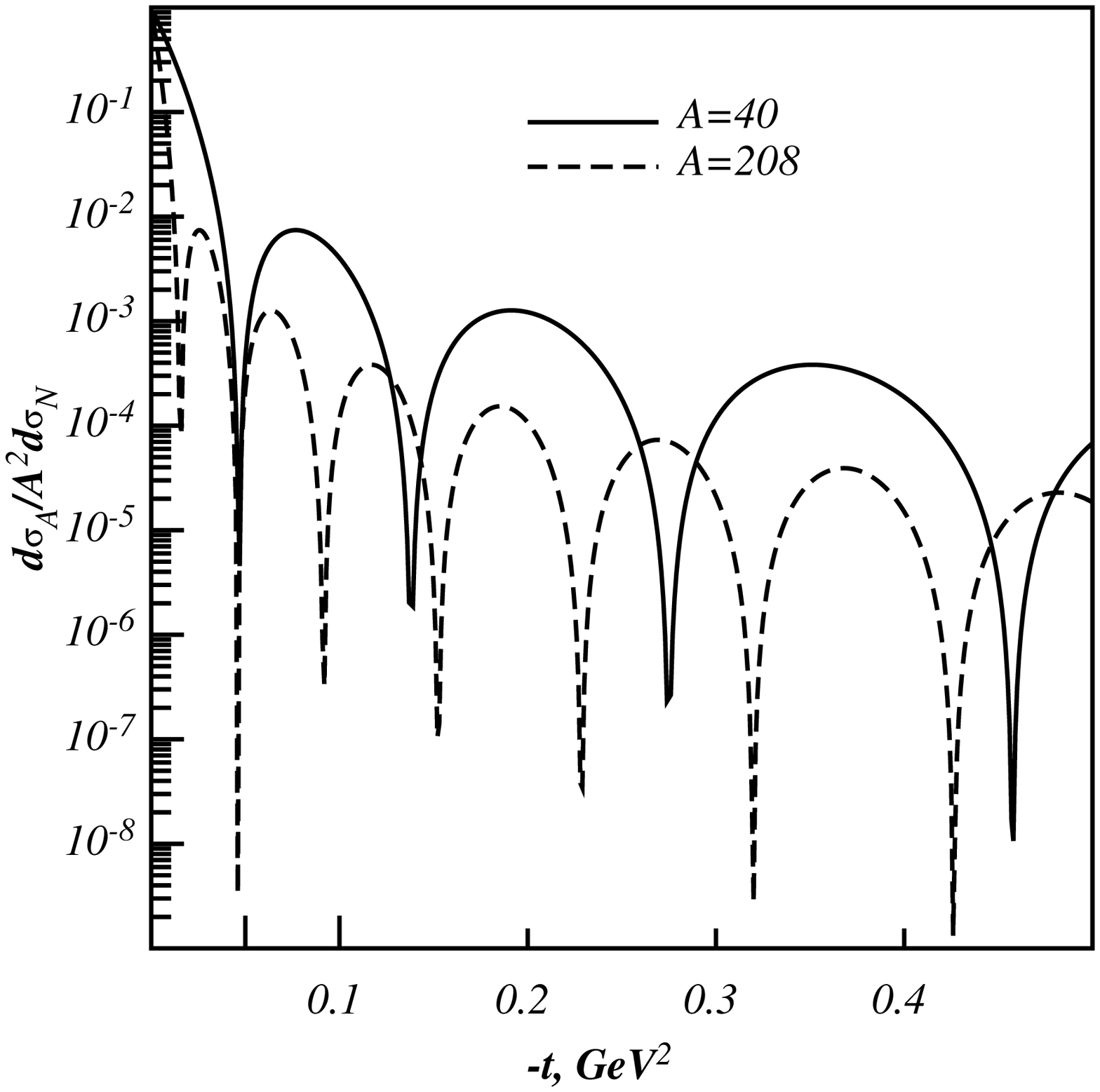}

\caption{\label{fig:ShadComparison} The nucleus to nucleon cross section ratio
for coherent DVCS as function of different kinematical variables.
Left: shadowing as function of Bjorken $x_{B}$ with and without gluon
shadowing for different nuclei, $t=t_{min},Q^{2}=5$GeV$^{2}$. Right:
$t$-dependence for different $A$. $x_{B}=10^{-3},\, Q^{2}=5$GeV$^{2}$.}

\end{figure}

In the Figure~\ref{fig:ShadComparison} we compare the results for
the coherent cross section ratio evaluated with and without  gluon
shadowing. As expected, gluon shadowing is very small at $x_{B}\gtrsim10^{-3}$
and gives $\sim20-30\%$ contribution for $x\sim10^{-5}$. Similar
dependence is observed for all the other cross-sections. In what follows,
for the sake of brevity we present only results which include gluon
shadowing.

\begin{figure}[h]
\includegraphics[scale=0.3]{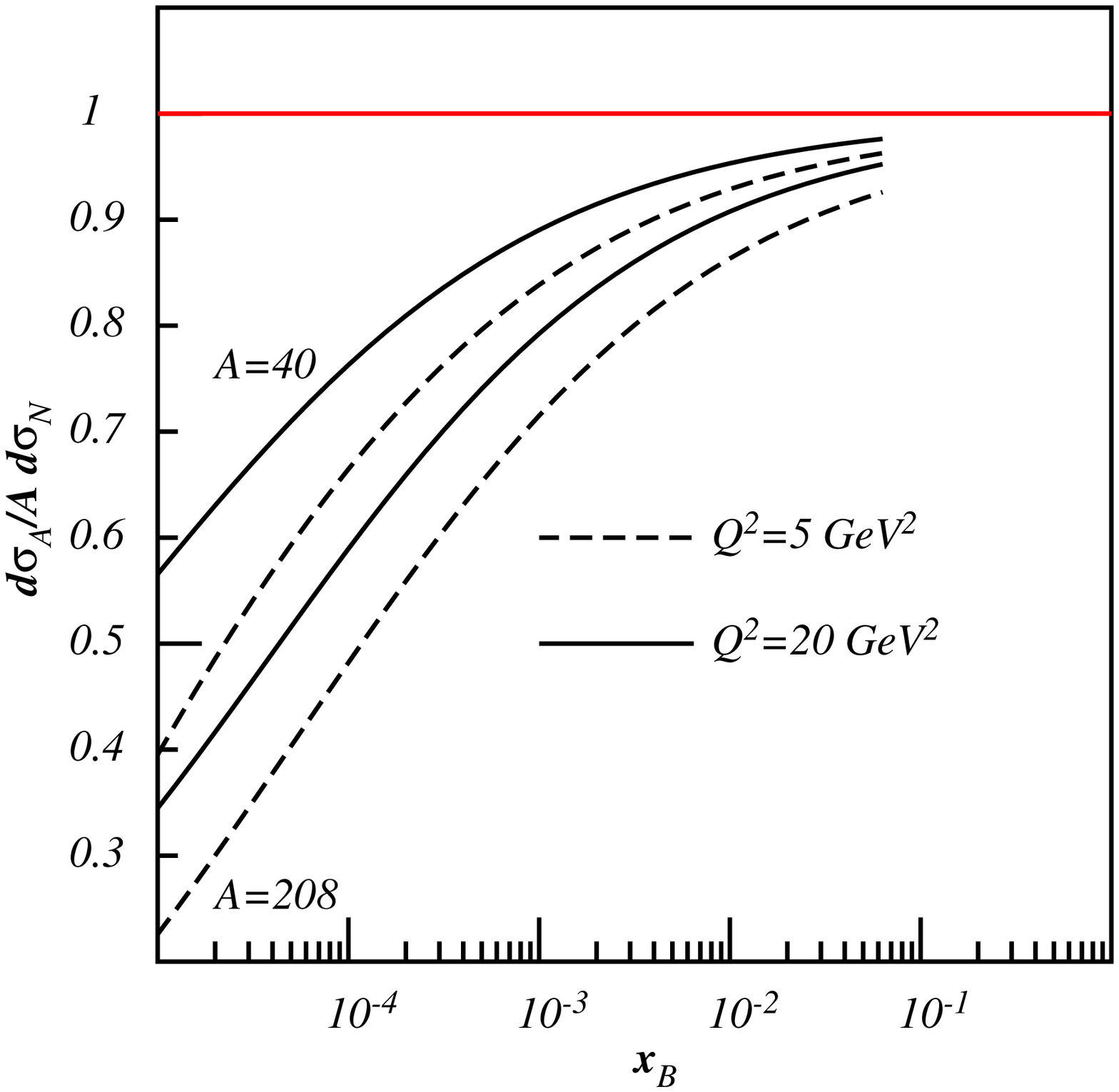}\includegraphics[scale=0.3]{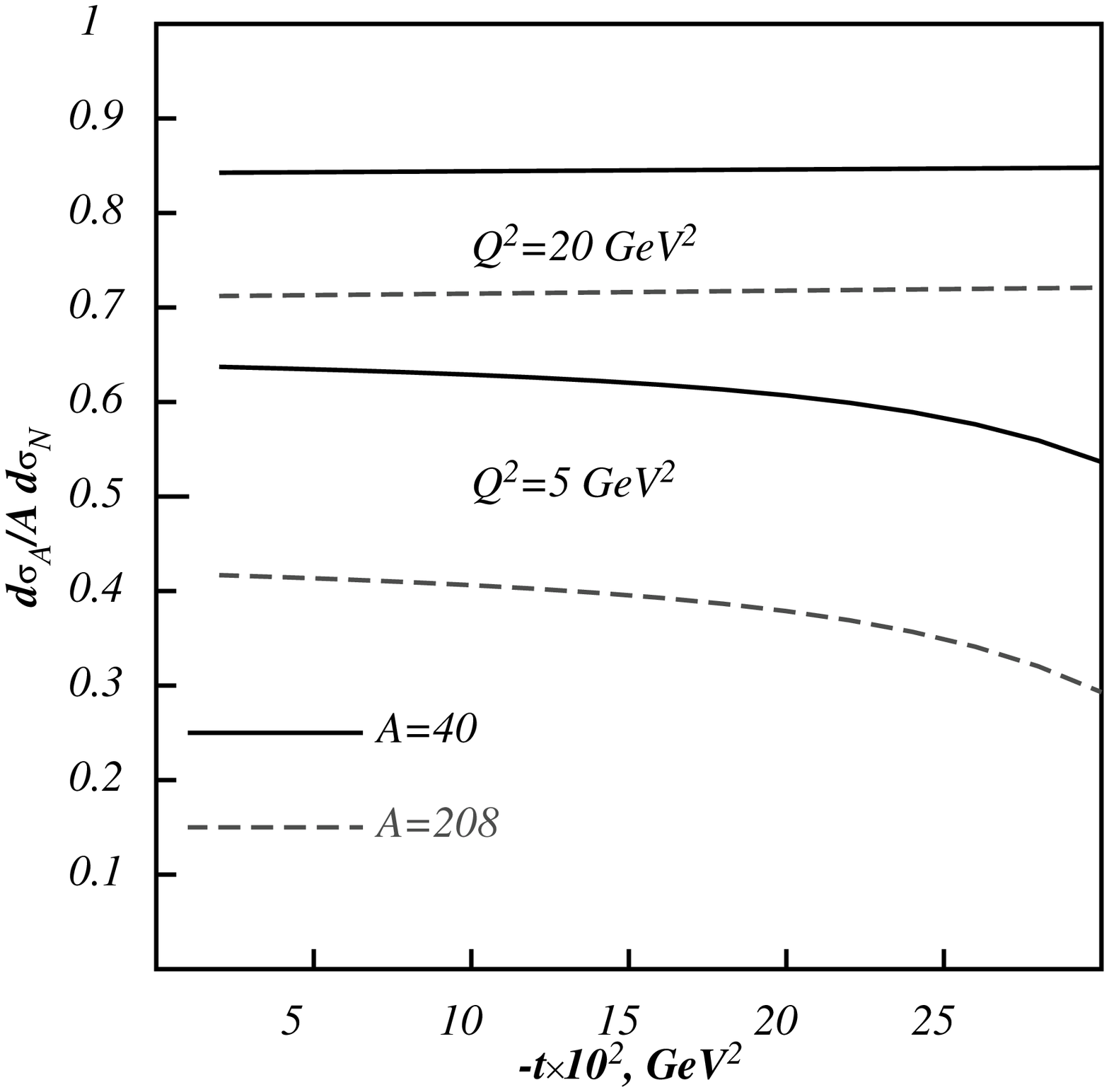}\includegraphics[scale=0.3]{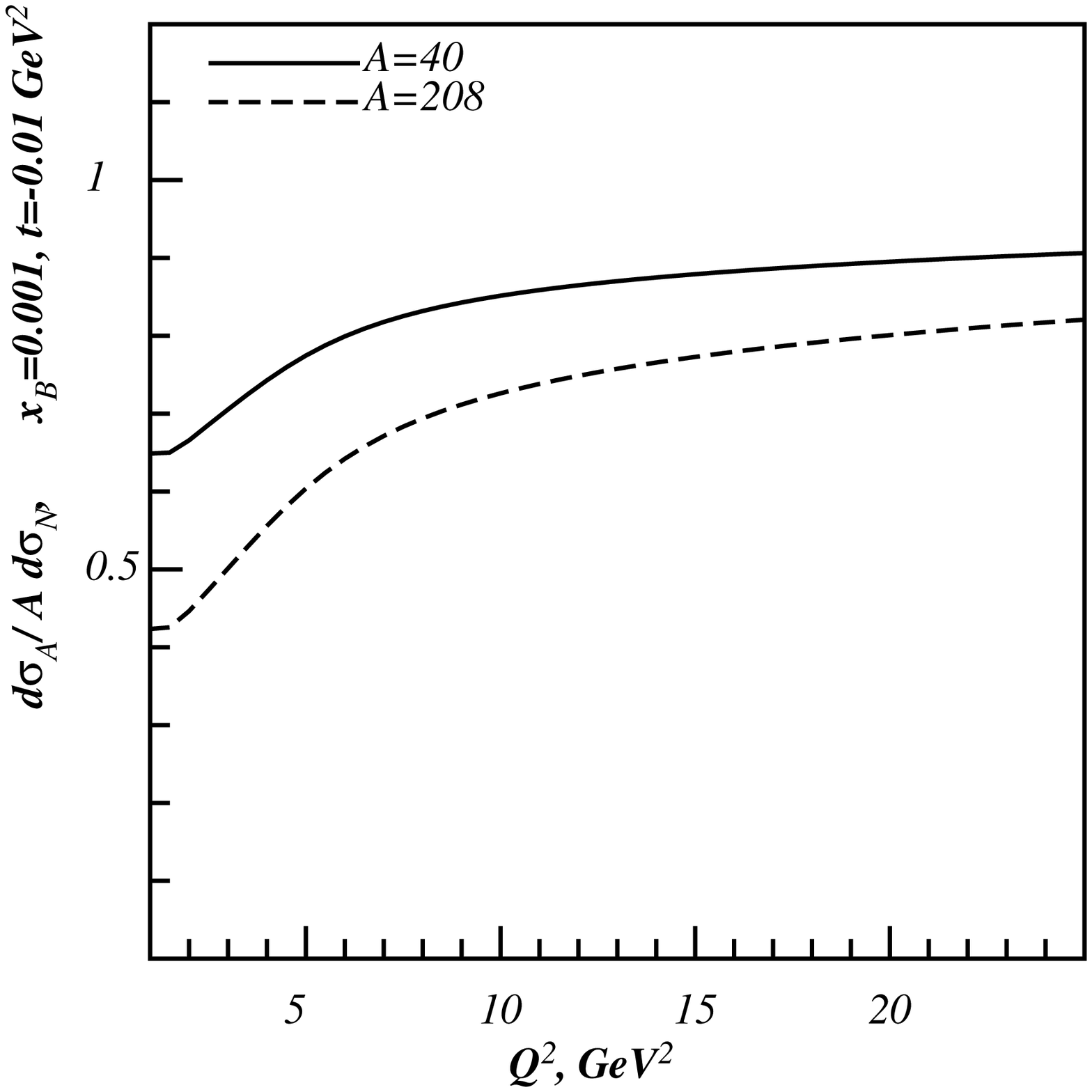}

\caption{\label{fig:DVCS-INCOH-RES}Dependence of the nuclear ratio for incoherent
DVCS on different kinematical variables. Left: $x_{B}$-dependence,
center: $t$-dependence, right: $Q^{2}$-dependence.}

\end{figure}

The nuclear ratio for incoherent scattering, $\gamma^{*}A\to\gamma A^{\prime}$,
is shown in the Figure~\ref{fig:DVCS-INCOH-RES}. It turns out that
the effect of shadowing at small $x_{B}$ is twice as strong as for
the coherent case. As a function of $t$, the nuclear cross-section
is almost a constant-due to the nucleus breakup the nuclear cross-section
is not suppressed by $F_{A}(t)$ and decreases only as a function
of the nucleon formfactor $F_{N}(t)$. The $Q^{2}$ dependence of
the nuclear ratio is quite similar to what was found for the coherent
case.

\begin{figure}[h]
\includegraphics[scale=0.3]{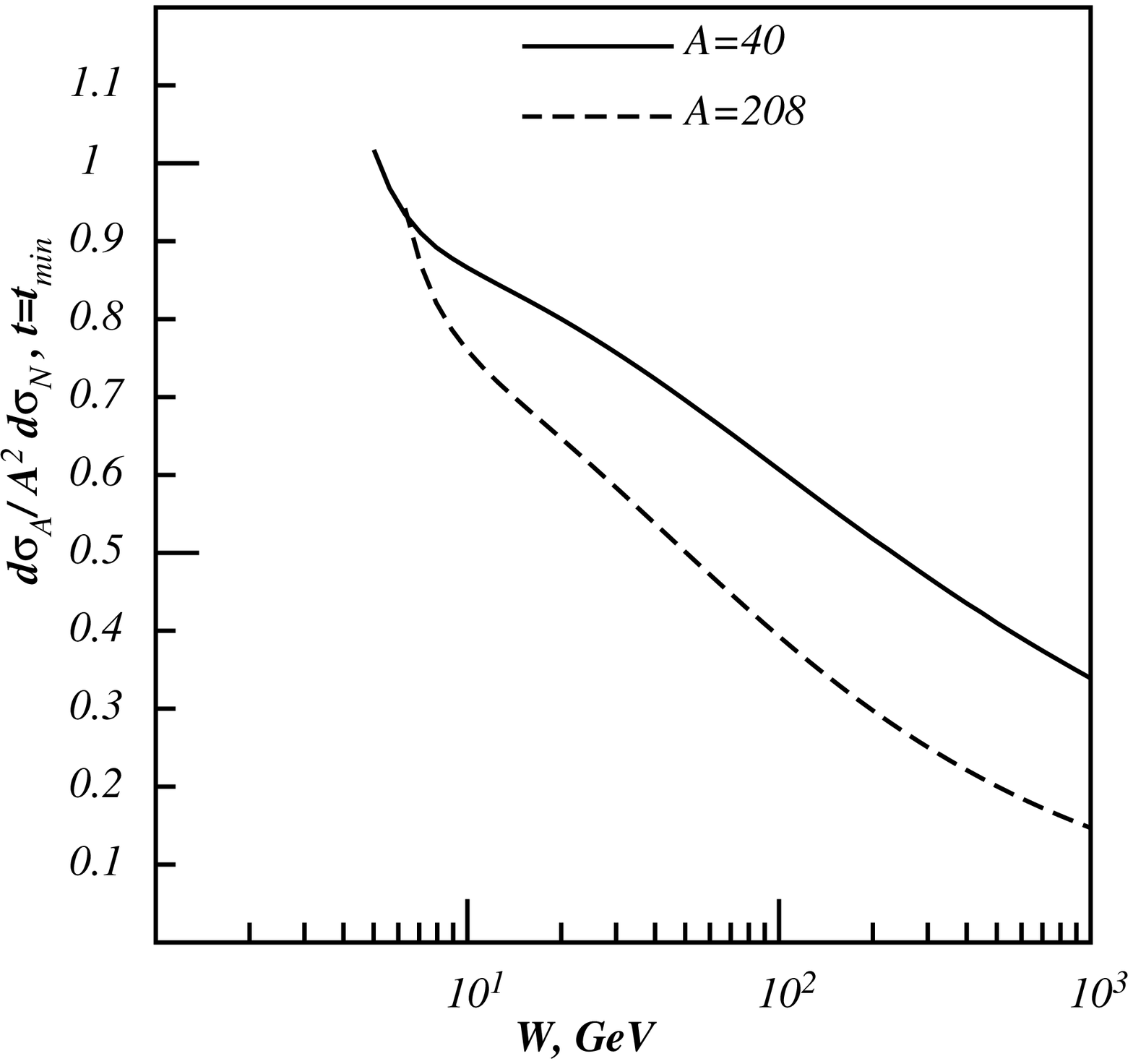} \includegraphics[scale=0.3]{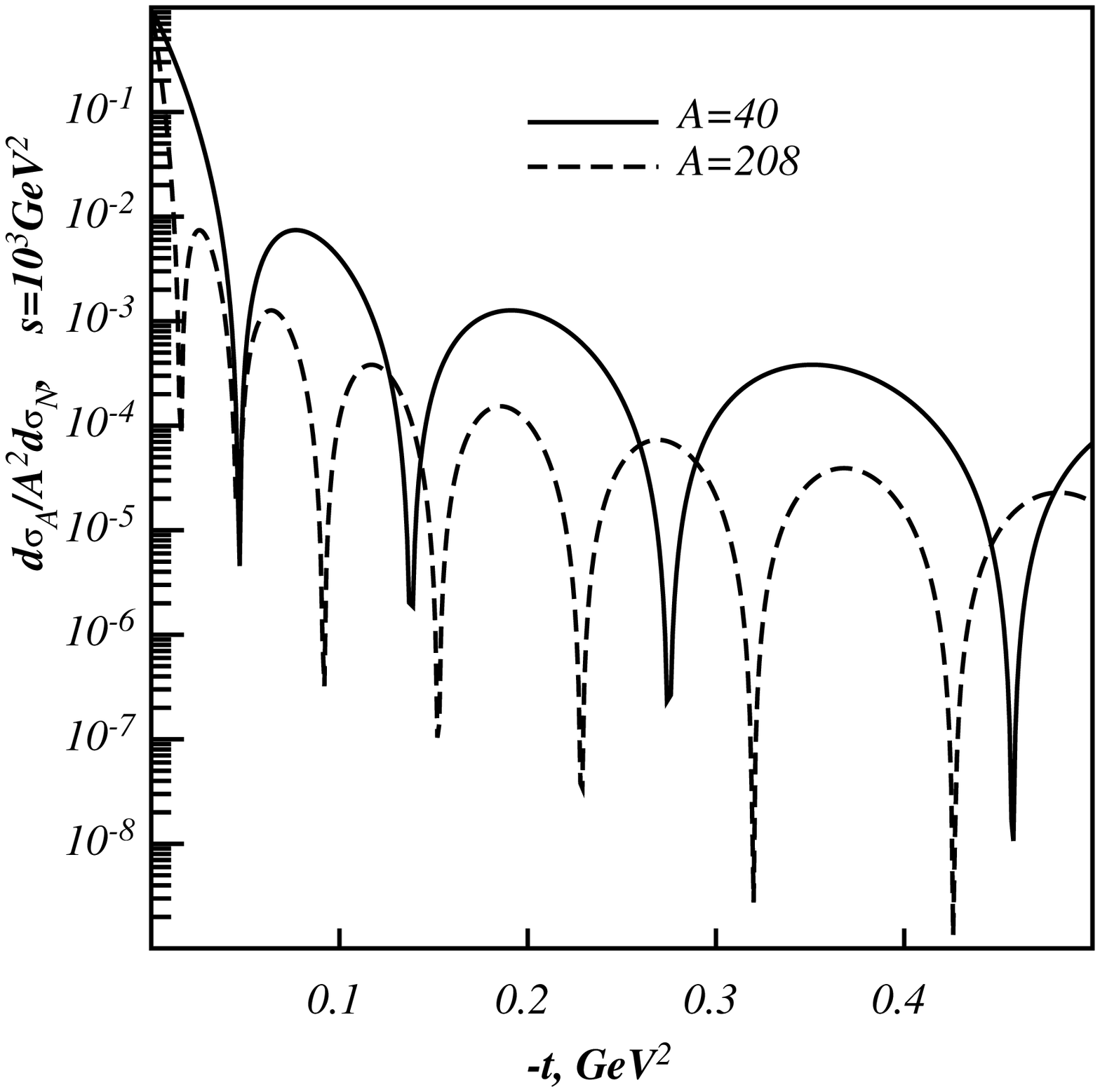}

\caption{\label{fig:RCS-COH-RES} Nuclear ratio for coherent RCS as function
of $W$ (left) and $t$ (right).}

\end{figure}

For coherent RCS, $\gamma A\to\gamma A$, the results of evaluation
are shown in Figure~\ref{fig:RCS-COH-RES}. Similar to DVCS, the
cross-section is steeply falling with center of mass energy $W$ (for
DVCS $x_{B}\propto1/W^{2}$ at fixed $\left(Q^{2},t\right)$). As
function of the momentum transfer $t$, the shadowing correction reveals
the behavior qualitatively similar to the case of coherent DVCS (see
Fig.~\ref{fig:DVCS-COH-RES}) It steeply decreases similar to the
behavior of the nuclear formfactor $F_{A}(t)$.

\begin{figure}[h]
\includegraphics[scale=0.3]{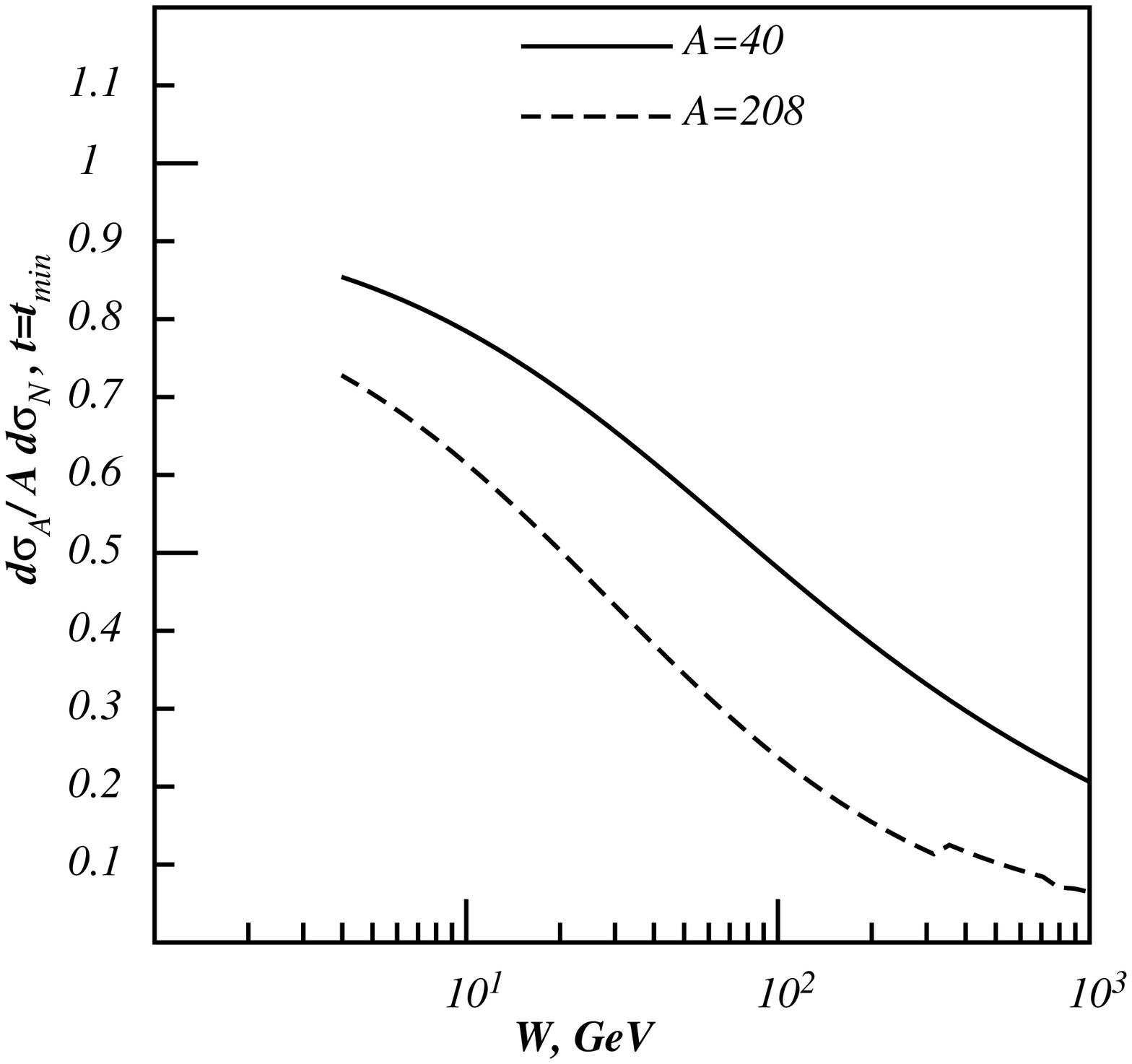} \includegraphics[scale=0.3]{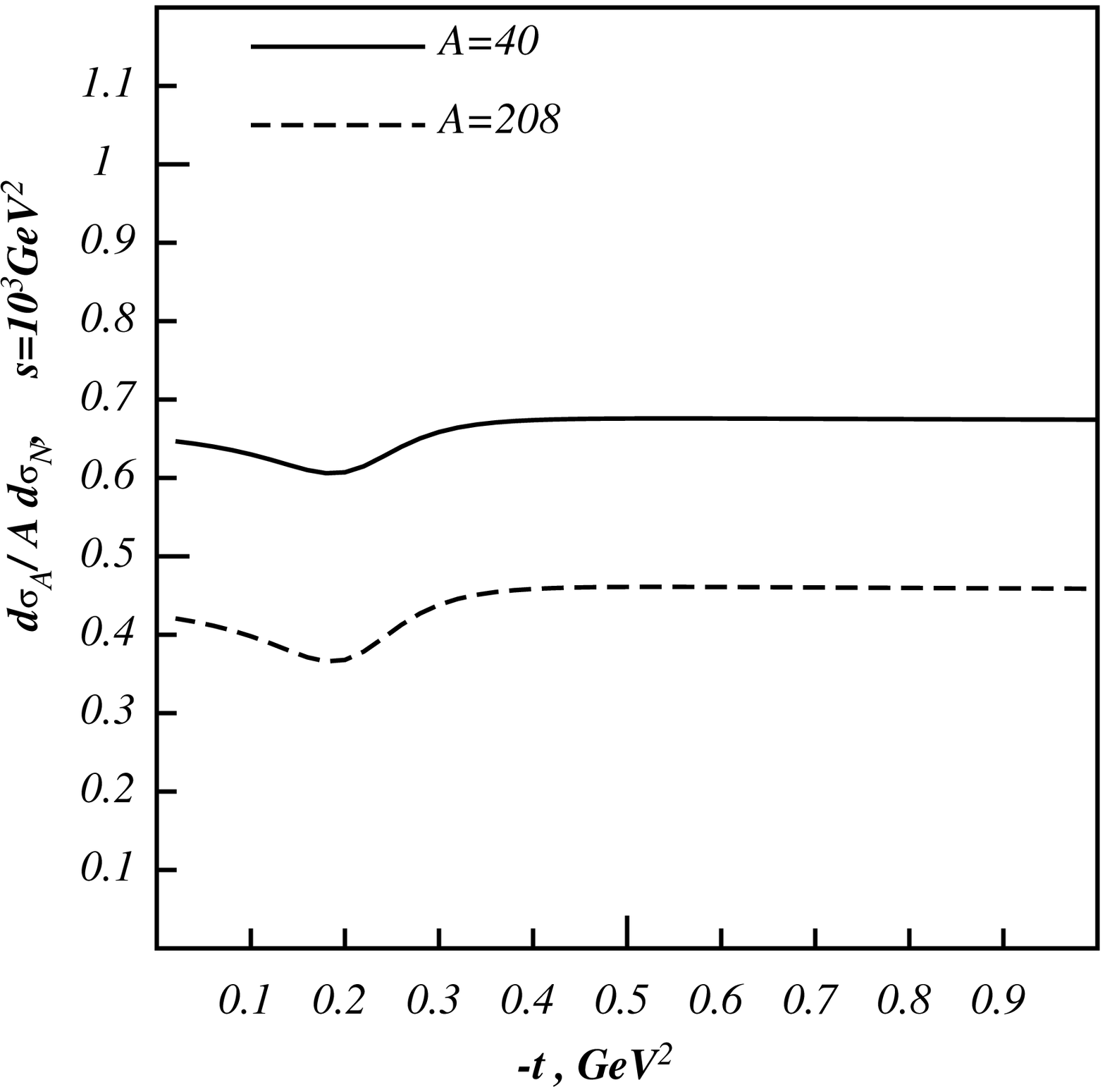}

\caption{\label{fig:RCS-INCOH-RES}Dependence of the incoherent nuclear RCS
cross-section ratio on $W$ (left) and $t$ (right).}

\end{figure}

For the incoherent RCS, $\gamma A\to\gamma A^{\prime}$, the results
of evaluation are shown in Figure~\ref{fig:RCS-INCOH-RES}. Similar
to the coherent case, the cross-section is decreasing as function
of $W$, down to the values an order of magnitude smaller than would
give a simple sum of the nucleon cross-sections. As function of $t$,
the cross-section is almost constant because of absence of the nuclear
formfactor.

\section{Conclusions and prospects}

\label{sec:Conclusions}

In this paper we considered DVCS and RCS on nuclear targets within
the color dipole model. Results for the coherent shadowing are presented
in Section~\ref{sec:Results}. We found that the magnitude of nuclear
shadowing is large and is important for analysis of DVCS and RCS data.
It may cause a substantial suppression of the nuclear DVCS cross-section.
We observed that for incoherent case, the shadowing is stronger than
for the coherent case. This happens since the incoherent cross section
is proportional to the survival probability of the dipole, so it vanishes
in the black disk limit ($\propto A^{1/3}$). At the same time, the
coherent cross section saturates at a value $\propto A^{2/3}$. For
the RCS, the shadowing is larger than for the DVCS, since a real photon
fluctuates to dipoles of a larger size than a virtual one.

Our results for DVCS and RCS can be applied to $eA$ collisions at
future electron-ion colliders (EIC and LHeC,~\cite{Klein:2008zza,Zimmermann:2008zzd}).
Currently data for nuclear DVCS are available from the HERMES experiment
at HERA~\cite{Airapetian:2009bi}, however the values of Bjorken
$x_{B}$ are too large to produce any sizable shadowing effects. Besides
electron beams, the usual source of quasi-real photons, one can~also
use beams of charged hadrons. Provided that the transverse overlap
of the colliding hadrons is small, i.e. the transverse distance $b$
between the centers of the colliding particles exceeds the sum of
their radii, $b>R_{1}+R_{2}$, the electromagnetic interaction between
colliding particles becomes the dominant mechanism. Such processes
called ultra-peripheral collisions (UPC) can be studied in $pp$,
$pA$ and $AA$ collisions. In particular, one can access RCS in the
reaction \begin{eqnarray}
A_{1}+A_{2} & \to A_{1}+\gamma+A_{2}.\label{eq:pp2ppgamma}\end{eqnarray}
 The typical virtualities $\left\langle Q_{\gamma^{*}}^{2}\right\rangle $
of the intermediate photon $\gamma^{*}$ are controlled by the formfactors
of the colliding particles, and are small: \begin{equation}
\left\langle Q_{\gamma^{*}}^{2}\right\rangle \lesssim\frac{3}{R_{A}^{2}}\sim\frac{0.1\, GeV^{2}}{A^{2/3}},\label{eq:Q2pp2ppgamma}\end{equation}
 where $A$ is the atomic number of the hadron ($A=1$ for a proton)
interacting electromagnetically, i.e. emitting the photon, while the
second hadron interacts strongly.

Thus, $\left\langle Q_{\gamma^{*}}^{2}\right\rangle $ is of the order
of the soft hadronic scale, so the intermediate photon can be treated
as a free Weizs�cker-Williams one, i.e. the amplitude of the process~(\ref{eq:pp2ppgamma})
can be described in terms of RCS.

These processes at the LHC will allow to study RCS at very high energies.
The possibility of observation of such processes experimentally has
been demonstrated by the STAR \cite{Abelev:2007nb,Adams:2004rz,Adler:2002sc}
and PHENIX~\cite{dEnterria:2006ep} experiments at RHIC. It is expected
that at LHC photon-proton collisions at energies up to $\sqrt{s_{\gamma p}}\lesssim8\times10^{3}$~GeV
can be observed~\cite{Baltz:2007kq}.

\section*{Acknowledgments}

This work was supported in part by Fondecyt (Chile) grants 1090291,
1090073, 1100287, and by DFG (Germany) grant PI182/3-1.

\end{document}